\documentclass{sf2a-conf2024}
\usepackage{graphicx}
\usepackage{hyperref}
\usepackage[]{natbib}  
\usepackage{epstopdf}

\def\BibTeX{{\rm B\kern-.05em{\sc i\kern-.025em b}\kern-.08em
    T\kern-.1667em\lower.7ex\hbox{E}\kern-.125emX}}
\bibpunct{(}{)}{;}{a}{}{,}  

\begin{document}

\TitreGlobal{SF2A 2024}


\title{Astro-COLIBRI: Empowering Citizen Scientists in Time Domain Astronomy}
\runningtitle{Astro-COLIBRI}

\author{Fabian Schüssler}\address{IRFU, CEA, Université Paris-Saclay, Gif-sur-Yvette, France}
\author{M. de Bony de Lavergne$^1$}
\author{A. Kaan Alkan$^{1,}$}\address{Laboratoire Interdisciplinaire des Sciences du Numérique, CNRS, Université Paris-Saclay, 91405 Orsay, France}
\author{J. Mourier$^1$}

\setcounter{page}{237}


\maketitle


\begin{abstract}
In recent decades, astronomy and astrophysics have experienced several fundamental changes. On one hand, there has been a significant increase in the observation of transient phenomena, which are short-lived events such as supernova explosions, fast radio bursts, and gamma-ray bursts. In addition, the detection of a growing number of different cosmic messengers provides researchers with crucial information about these objects. For example, the detection of high-energy neutrinos and gravitational waves regularly complements traditional astronomical observations in the electromagnetic spectrum. This trend is expected to intensify in the coming years with the commissioning of a wide variety of next-generation observatories, which will enable more in-depth studies of the transient sky.

To enhance our understanding and optimize the observations of these phenomena, we have developed the Astro-COLIBRI platform. It is freely available to amateur and professional astronomers in the form of a smartphone application (iOS and Android), a web interface (\url{https://astro-colibri.com}), an API (\url{https://astro-colibri.science}), and a chatbot 'Astro-COLIBRI GPT'\footnotetext{\url{https://gptstore.ai/gpts/sQHXeqV9cE-astro-colibri}}, among many other features. Astro-COLIBRI serves as a central access point for information on astrophysical sources and transient events, allowing a wide network of observers to track and receive real-time alerts.

Here we highlight the key features of Astro-COLIBRI, with a particular emphasis on recent innovations. These include a discussion forum that facilitates user interactions and our strengthened collaboration with various networks of amateur astronomers. \end{abstract}



\section{Introduction}
Astro-COLIBRI~\citep{2021ApJS..256....5R, 2023Galax..11...22R} is a novel platform designed to enhance the study of transient astronomical events by integrating real-time multi-messenger observation tools into a user-friendly interface. It serves both professional and amateur astronomers, providing a comprehensive solution for coordinating follow-up observations and maximizing the discovery potential of transient phenomena. By consolidating alerts and data from all publicly accessible observatories and distribution systems, Astro-COLIBRI facilitates the rapid identification and analysis of events such as Gamma-Ray Bursts (GRBs), Fast Radio Bursts (FRBs), supernovae, and gravitational waves.

The primary aim of Astro-COLIBRI is to streamline the process of observing these events, which are unpredictable, often rapidly fading, and thus require immediate attention. By offering real-time alerts and detailed information on each new detection, the platform enables astronomers to respond swiftly and efficiently to new discoveries. The integration of multi-wavelength and multi-messenger data allows for a comprehensive understanding of these events, enhancing the ability to conduct in-depth studies and collaborative efforts.

Astro-COLIBRI's architecture includes a public RESTful API, real-time databases, a cloud-based alert system, and both web and mobile applications. This infrastructure supports a wide range of astrophysical phenomena and provides users with tools to plan and execute observations across different wavelengths and messengers. The platform's user-friendly graphical interfaces, available on both iOS and Android, make it accessible to a broad audience, including citizen scientists who play an increasingly important role in the global observation network.

\section{Purpose and Core Features}

\begin{figure}[t!]
\centering
     \begin{minipage}{0.45\textwidth}
        \centering
        \includegraphics[width=\textwidth]{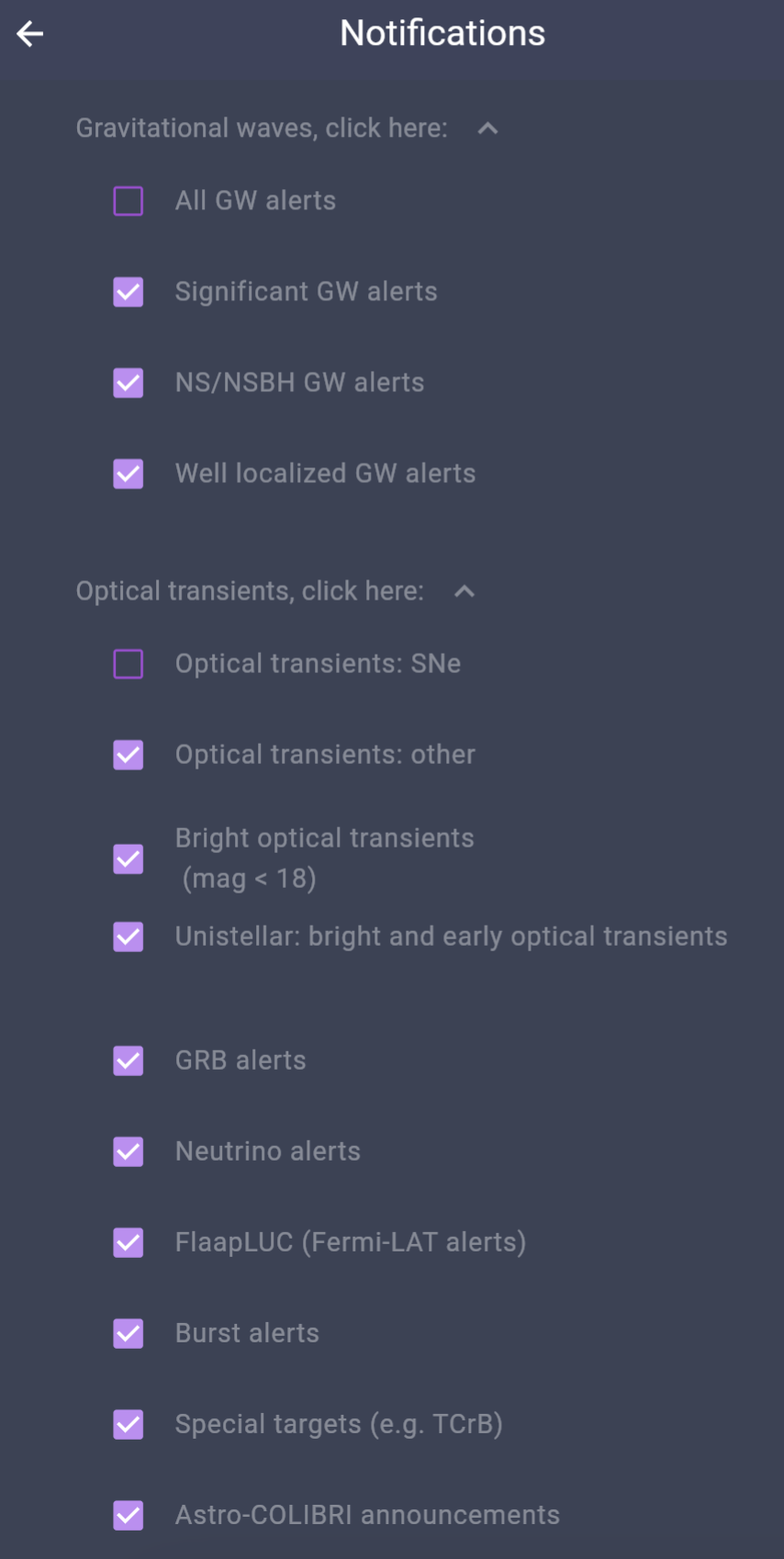}
\caption{Selection of realtime notification streams.}
\label{fig:notifications}
    \end{minipage} \hfill
    \begin{minipage}{0.45\textwidth}
        \centering
        \includegraphics[width=\textwidth]{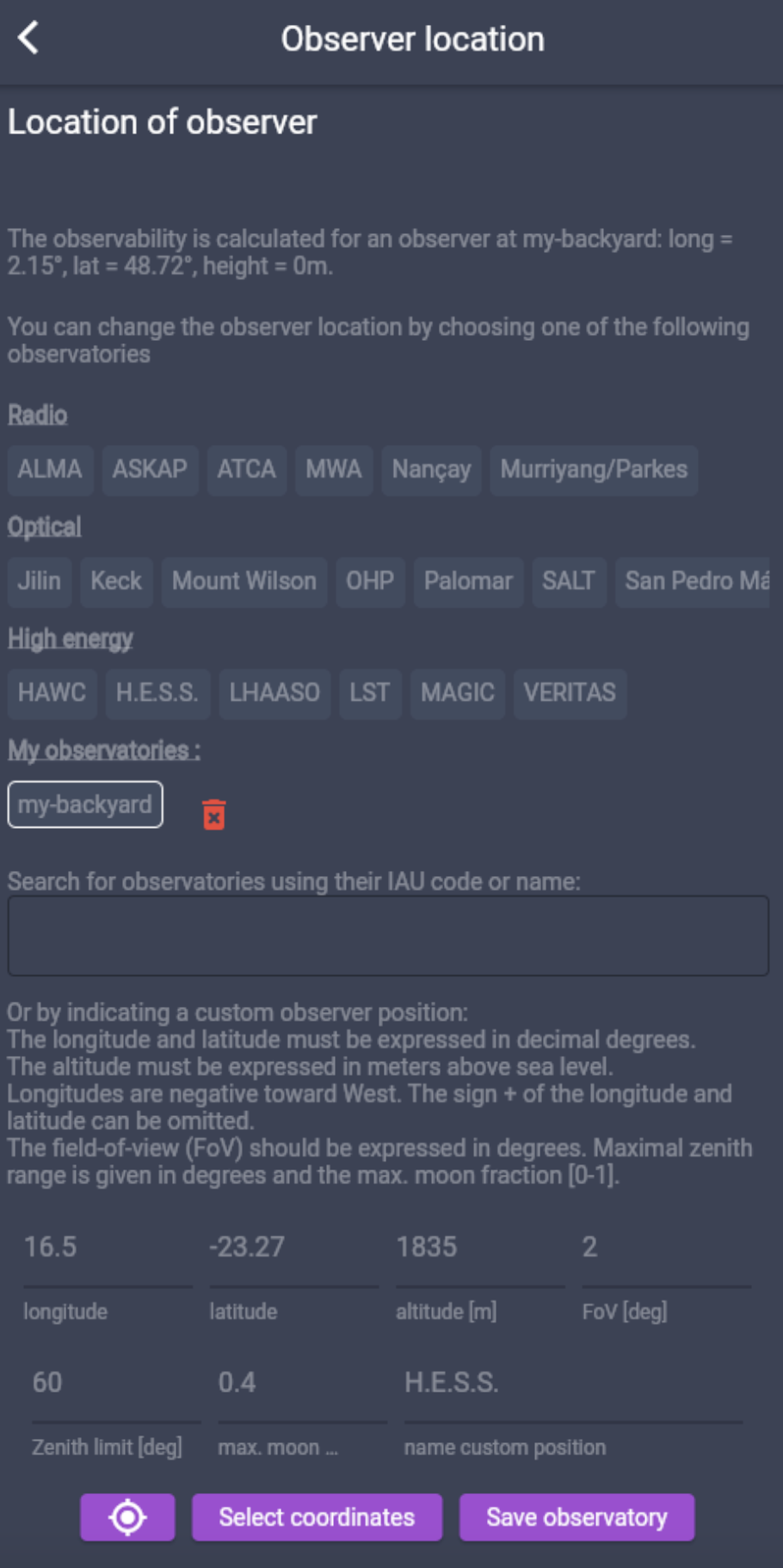} 
        \caption{Selection of follow-up observatories.}\label{fig:observatory}
    \end{minipage}
\end{figure}

Astro-COLIBRI was developed to address the need for a unified system that can handle the complex requirements of observations of transient phenomena. The platform's core features are designed to provide comprehensive support for multi-messenger astrophysics:

\begin{description}
    \item[Real-Time Alerts and Notifications:] Astro-COLIBRI evaluates incoming messages from various alert streams in real-time, filtering them based on user-specified criteria. This ensures that astronomers receive timely notifications about relevant events, enabling quick decision-making and follow-up actions. An overview of the available notification streams is given in Fig.~\ref{fig:notifications}.

\item[Graphical User Interfaces:] The platform offers intuitive web and mobile interfaces that display an overview of recent transient events, including their locations, classifications, and relevant observational data. Users can customize the display settings and filter options to suit their specific needs.

\item[Observatory Selection and Scheduling:] Astro-COLIBRI includes tools for selecting observatories and scheduling observations. Users can define custom observer locations, including professional and amateur observatories, and generate optimized observation plans for events such as gravitational wave detections~\citep{tilepy_icrc2023}.

\item[Multi-Wavelength and Multi-Messenger Integration:] The platform integrates data from various sources, including high-energy neutrinos, optical transients, gamma-ray bursts, and gravitational waves. This allows for a comprehensive overview of events and enhances the potential for new discoveries through coordinated multi-messenger observations.

\item[API Access:] The public RESTful API enables the integration of Astro-COLIBRI's data into external environments and tools. This flexibility supports a wide range of use cases, from professional observatories to amateur astronomer setups, facilitating broader participation in transient event studies.

\item[Citizen Science Support:] Recognizing the valuable contributions of amateur astronomers, Astro-COLIBRI provides features tailored to their needs, such as notifications for bright events and tools for defining custom observatory locations. This encourages greater involvement from the citizen science community and enhances the overall observational network.

\end{description}


\section{Observatory selection}
A crucial aspect of follow-up observations for a particular event is determining its observability from a specific location. This can be specified via the "Observer location" option found in the app menu at the top right corner or on the web interface by clicking the teardrop icon in the "personalize" section in the top row. As shown in Fig.~\ref{fig:observatory}, there are three complementary methods to select the location of the observer:

\begin{description}
    \item[Pre-defined observatories:] A collection of professional observatories worldwide, categorized by wavelength range, is available for quick selection.
    \item[Searches in the IAU database:] Users can search the International Astronomical Union (IAU) database for additional observatories using an autocomplete-enabled text field.
    \item[Custom observatories:] Ideal for citizen scientists and amateur astronomers, custom locations can be specified. Users can either use the GPS position of their device or manually input the coordinates and parameters of their setup. To save these entries for future use, a user account (free to create with a valid email address) is required.
\end{description}

\section{Time domain observations by amateur astronomers}
Astro-COLIBRI provides free and easy access to astrophysical time-domain observations that were previously only available to professional astronomers due to various, mainly technical, barriers. Interested amateur astronomers and citizen scientists are now obtaining the same information used by professional observatories around the world. They can thus effectively conduct follow-up observations and fully participate in the scientific process. Several collaborative tools and dedicated features are available to facilitate this process.

\subsection{Dedicated notifications for Bright Events}
Astro-COLIBRI provides specialized notifications for bright, optical sources that are within the reach of amateur astronomers' instrumentation. These notifications highlight phenomena such as supernovae, novae, tidal disruption events, and other optical transients with magnitudes suitable for observation with small to medium-sized telescopes (i.e. $\mathrm{mag} < 18$). This ensures that amateur astronomers receive timely alerts about events they can realistically observe and contribute valuable data to the astronomical community.

\subsection{Detailed event selection}
The platform offers detailed event selection capabilities, allowing users to filter and prioritize the most relevant transient events based on their preferences and observational capabilities. This feature helps amateur astronomers focus their efforts on the most promising targets, optimizing their observation schedules, and increasing the likelihood of significant findings.

\subsection{Integration with the RAPAS and BHTOM networks}

Members of the RAPAS~\citep{RAPAS} and BHTOM~\citep{BHTOM} networks benefit from the ability to perform manual event selections and submit them directly from the Astro-COLIBRI interfaces. This seamless integration facilitates coordinated observation campaigns and data sharing among network members, enhancing the overall impact of their contributions. The platform's user-friendly interface simplifies the process of identifying, selecting, and sharing observations within the networks and with the broader astronomical community.

\subsection{Weekly List of Top Transient Events}

In collaboration with the RAPAS network, Astro-COLIBRI compiles each week a list of the ten most interesting transient events and posts it to a dedicated forum\footnotemark
\footnotetext{\url{https://forum.astro-colibri.science}\label{fn:forum}}. This curated list helps amateur astronomers stay informed about the latest and most significant events, providing a valuable resource for planning their observations. The weekly list is carefully selected to include events that offer high scientific potential and are accessible to a wide range of observers.

\subsection{Discussion Forum and Community Involvement}

The Astro-COLIBRI discussion forum\footnotemark[\value{footnote}] serves as a vibrant hub for interaction between amateur and professional astronomers. Users can discuss interesting events, share observations, and exchange insights about the latest transient phenomena. The forum also features a voting system that allows members to vote for the most interesting events on the weekly Top-10 list to observe. This collaborative environment fosters a sense of community and encourages collective decision-making, ultimately enhancing the effectiveness of observational efforts.

\section{Conclusions}
Astro-COLIBRI represents a significant advancement in the field of multi-messenger astrophysics, offering a comprehensive and efficient solution for real-time observation and analysis of transient astronomical events. By integrating various data sources and providing user-friendly interfaces, the platform enables both professional and amateur astronomers to better understand and explore the dynamic universe. The collaborative potential facilitated by Astro-COLIBRI enhances the overall scientific output and fosters a more connected and informed astronomical community.

The Astro-COLIBRI development team welcomes comments and feedback from the community to further improve the platform. Everybody is invited to join our discussion forum\footnotemark[\value{footnote}] or contact the developer team at \href{mailto:astro.colibri@gmail.com}{astro.colibri@gmail.com}.

\section{Acknowledgements}
The authors acknowledge the support of the French Agence Nationale de la Recherche (ANR) under reference ANR-22-CE31-0012. This work was also supported by the Programme National des Hautes Energies of CNRS/INSU with INP and IN2P3, co-funded by CEA and CNES and we acknowledge support by the European Union’s Horizon 2020 Programme under the AHEAD2020 project (grant agreement n. 871158).

\bibliographystyle{aa}  
\bibliography{SCHUSSLER_S21} 

\begin{thebibliography}{5}
\expandafter\ifx\csname natexlab\endcsname\relax\def\natexlab#1{#1}\fi

\bibitem[{{Reichherzer, P., Sch{\"u}ssler, F., Lefranc, V., Becker Tjus, J., Mourier, J., Kaan Alkan, A.}(2023)}]{2023Galax..11...22R}
{Reichherzer, P., Sch{\"u}ssler, F., Lefranc, V., Becker Tjus, J., Mourier, J., Kaan Alkan, A.} 2023, Galaxies, 11, 22

\bibitem[{{Reichherzer, P., Sch{\"u}ssler, F., Lefranc, V., Yusafzai, A., Kaan Alkan, A., Ashkar, H., Becker-Tjus, J.}(2021)}]{2021ApJS..256....5R}
{Reichherzer, P., Sch{\"u}ssler, F., Lefranc, V., Yusafzai, A., Kaan Alkan, A., Ashkar, H., Becker-Tjus, J.} 2021, ApJS, 256, 5

\bibitem[{{Sch\"ussler} {et~al.}(2023){Sch\"ussler}, {Ashkar}, {de Bony de Lavergne}, \& {Seglar-Arroyo}}]{tilepy_icrc2023}
{Sch\"ussler}, F., {Ashkar}, H., {de Bony de Lavergne}, M., \& {Seglar-Arroyo}, M. 2023, in 38th International Cosmic Ray Conference. July 26 - Aug 3, 2023. Nagoya, PoS(ICRC2023)1469

\bibitem[{{Thuillot} {et~al.}(2024){Thuillot}, {Midavaine}, M, {Buill}, \& {Neveu}}]{RAPAS}
{Thuillot}, W., {Midavaine}, T., M, D., {Buill}, C., \& {Neveu}, S. 2024, RAPAS, \url{https://rapas.imcce.fr/}

\bibitem[{{Wyrzykowski} {et~al.}(2024)}]{BHTOM}
{Wyrzykowski}, L. {et~al.} 2024, BHTOM, \url{https://bh-tom2.astrolabs.pl/}

\end{thebibliography}

\end{document}